\documentclass[
    journal,
    twocolumn,
    ]{IEEEtranTCOM}
\usepackage{graphicx}
\usepackage{amsmath}
\usepackage{amssymb}
\usepackage{amsthm}
\usepackage{afterpage}
\usepackage{verbatim}
\usepackage{psfrag}
\usepackage{color}
\usepackage{cite}
\usepackage{setspace}
\usepackage{epsfig}
\usepackage{epstopdf}
\usepackage{footmisc}
\usepackage{fmtcount}
\usepackage{stackrel}
\usepackage{float}
\usepackage{breqn}
\usepackage{enumerate}
\usepackage{graphicx}
\usepackage{subcaption}
\usepackage{graphicx}

\begin{document}
\title{On Achievable Accuracy of Localization in Magnetic Induction-Based Internet of Underground Things for Oil and Gas Reservoirs}
\author{Nasir Saeed,~\IEEEmembership{Member,~IEEE}, Mohamed-Slim Alouini,~\IEEEmembership{Fellow,~IEEE}, Tareq Y. Al-Naffouri,~\IEEEmembership{Member,~IEEE} 
\thanks{This work is
supported by Office of Sponsored Research (OSR) at King Abdullah University of Science and Technology (KAUST). 

The authors are with the Computer Electrical and Mathematical Sciences \& Engineering (CEMSE) Division, KAUST, Thuwal, Makkah Province, Kingdom of Saudi Arabia, 23955-6900.}
}
\maketitle
\begin{abstract}
Magnetic Induction (MI) is an efficient wireless communication method to deploy operational internet of underground things (IOUT) for oil and gas reservoirs. The IOUT consists of underground things which are capable of sensing the underground environment  and communicating with the surface. The MI-based IOUT enable many applications, such as monitoring of the oil rigs, optimized fracturing, and optimized extraction. Most of these applications are dependent on the location of the underground things and therefore require accurate localization techniques. The existing localization techniques for MI-based underground sensing networks are two-dimensional and do not characterize the achievable accuracy of the developed methods which are both crucial and challenging tasks. Therefore, this paper presents the expression of the Cramer Rao lower bound (CRLB) for three-dimensional MI-based IOUT localization which takes into account the channel parameters of the underground magnetic-induction.  The derived CRLB provide the suggestions for an MI-based underground localization system by associating the system parameters with the error trend. Numerical results demonstrate that localization accuracy is affected by different channel and networks parameters such as the number of anchors, noise variance, frequency, and the number of underground things.
\end{abstract}
\IEEEpeerreviewmaketitle

\section{Introduction}
\begin{figure}
\centering
\includegraphics[width=1\columnwidth]{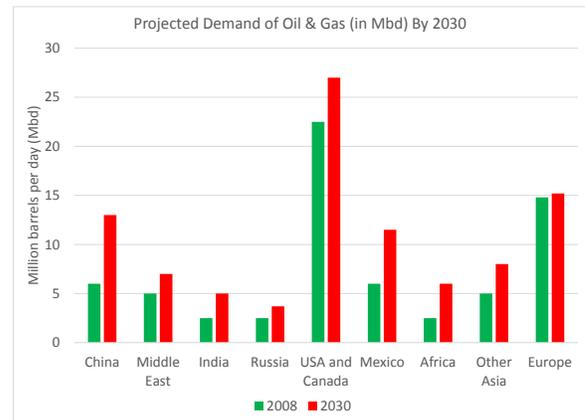}  
\caption{World projected energy demand by IEA.\label{fig:chart}} 
\vspace{-1.5 em}
\end{figure}

According to the report of the international energy agency (IEA), the energy needs of the world is expected to escalate by 40 \% in 2030 (see Fig.~\ref{fig:chart}) \cite{alcatel2008}. This ever-increasing demand for energy constitutes 20 \% and 50 \% escalation from the oil and gas industries respectively. However, the operating environment of the oil and gas industries is challenging to fulfill this inexorable demand for energy \cite{William2016}. One of the primary challenges of the underground oil and gas reservoirs is to obtain their real-time information. This challenge can be addressed by using internet of underground things (IOUT) which can optimize the production of oil and gas, monitor the flow of oil and gas, and can monitor the reservoir \cite{Kisseleff2018}. The IOUT based intelligent oil and gas fields can improve the accuracy, integrity, and timeliness of the production process \cite{Chen2014}. 

Although IOUT enable multiple applications for underground oil and gas reservoirs, the challenging underground environment prevents the use of conventional terrestrial wireless communication systems. Therefore, magnetic-induction (MI) has emerged as a promising wireless communication technology to develop practical underground sensing systems \cite{Sojdehei2001, Bansal2004, Akyildiz2006, Sun2013, Tan2015, Charaf2017}. MI uses time-varying magnetic fields to transmit the information in the underground environment. MI provide reliable and long-range communication in the underground environment as compared to its counterpart electromagnetic waves (EM) \cite{Abrudan2016}. The performance of the EM in the underground is profoundly affected by properties of the subsurface environment such as soil structure, underground medium (soil, water, sand, etc.), and water contents \cite{Franconi2014}. However, all these impediments cause less attenuation to the MI-based communication systems. Besides, the EM-based systems require large size antennas for communication which is impractical for the underground environment. Hence, MI-based systems are more practical because it rely on tiny size coil antennas. 

Subsequently, efforts have been made in the recent past to develop MI-based underground sensing systems. However, most of the applications of the underground sensing systems such as monitoring of oil rig, optimized fracturing, and collecting of geo-tagged sensing data, require location information of the deployed sensors (underground things) \cite{Abrudan2016, Lin2017}. Therefore, the authors in \cite{Lin2017} have proposed a localization scheme for underground sensing which utilizes the magnetic induction channel for distance estimation. The authors have used semi-definite programming (SDP) relaxation method for the distance estimation whereas the sensor nodes position is estimated by leveraging alternating direction augmented Lagrangian method (ADM) and conjugate gradient (CG) technique. The proposed solution in \cite{Lin2017} is for two-dimensional underground sensing network while the three-dimensional (3D) nature of the underground environment requires 3D localization which is more challenging. Moreover, the existing localization solutions do not consider the achievable accuracy for MI-based underground sensing systems which is a crucial design parameter for any positioning system. Therefore, motivated by the above challenges, we present in this article the theoretical accuracy limits for three-dimensional (3D) MI-based IOUT. The major contributions of the paper are summarized as follows:
\begin{itemize}
\item A realistic 3D architecture of MI-based IOUT is presented for oil and gas reservoirs.
\item A closed form expression for the Cramer Rao Lower bound (CRLB) is derived for 3D MI-based IOUT localization. The derived lower bound is useful to compare the results of different MI-based underground localization systems.
\item Numerical results to evaluate the performance of the derived localization bound concerning different channel and network parameters such as range measurement errors, network size, and number of anchors for MI-based IOUT. 
\end{itemize}

The remainder of the paper is organized as follows. In section II and III, we present the related work and system model respectively. Section IV introduces the formulation of the CRLB for MI-based IOUT localization. In section V, we provide numerical results to validate the theoretical analysis. Finally, section VI concludes the paper.

\section{Related Work}\label{sec:related}

The literature on localization techniques for terrestrial and underwater wireless networks is rich. In \cite{Paul2017} and \cite{Saeedsurvey2018} classification of localization techniques for terrestrial and marine wireless networks is presented respectively where the localization schemes are categorized based on the type of computation (centralized/distributed), ranging technique (range-based/range-free), and space (2D/3D). However, the research work on the development of localization systems for the underground environment is limited due to numerous challenges such as non-availability of global positioning system (GPS) signals, high attenuation of radio frequency (RF) and electromagnetic (EM) waves, light-less environment, and narrow operational area. Even though localization techniques for GPS-denied environment such as underwater or indoor have been well developed, it is hard to apply those techniques to underground localization \cite{Kisseleff2018}. Both underwater and indoor localization techniques are based on either RF, acoustic, or optical signals. However, the subsurface environment does not support the use of these signals, and therefore the localization techniques developed for the underwater or indoor environment cannot be directly applied. Thus, a two-dimensional (2D) localization technique has been proposed in \cite{Lin2017} for MI-based sensing networks.  The authors in \cite{Lin2017} have introduced the use of MI induction for channel based distance estimation where the underground sensors were able to estimate the distances to their neighbors and the anchor nodes. Furthermore, a modified SDP relaxation based technique is used which jointly uses ADM and CG to determine the final position of the sensors. 
\begin{figure}
\centering
\includegraphics[width=1\columnwidth]{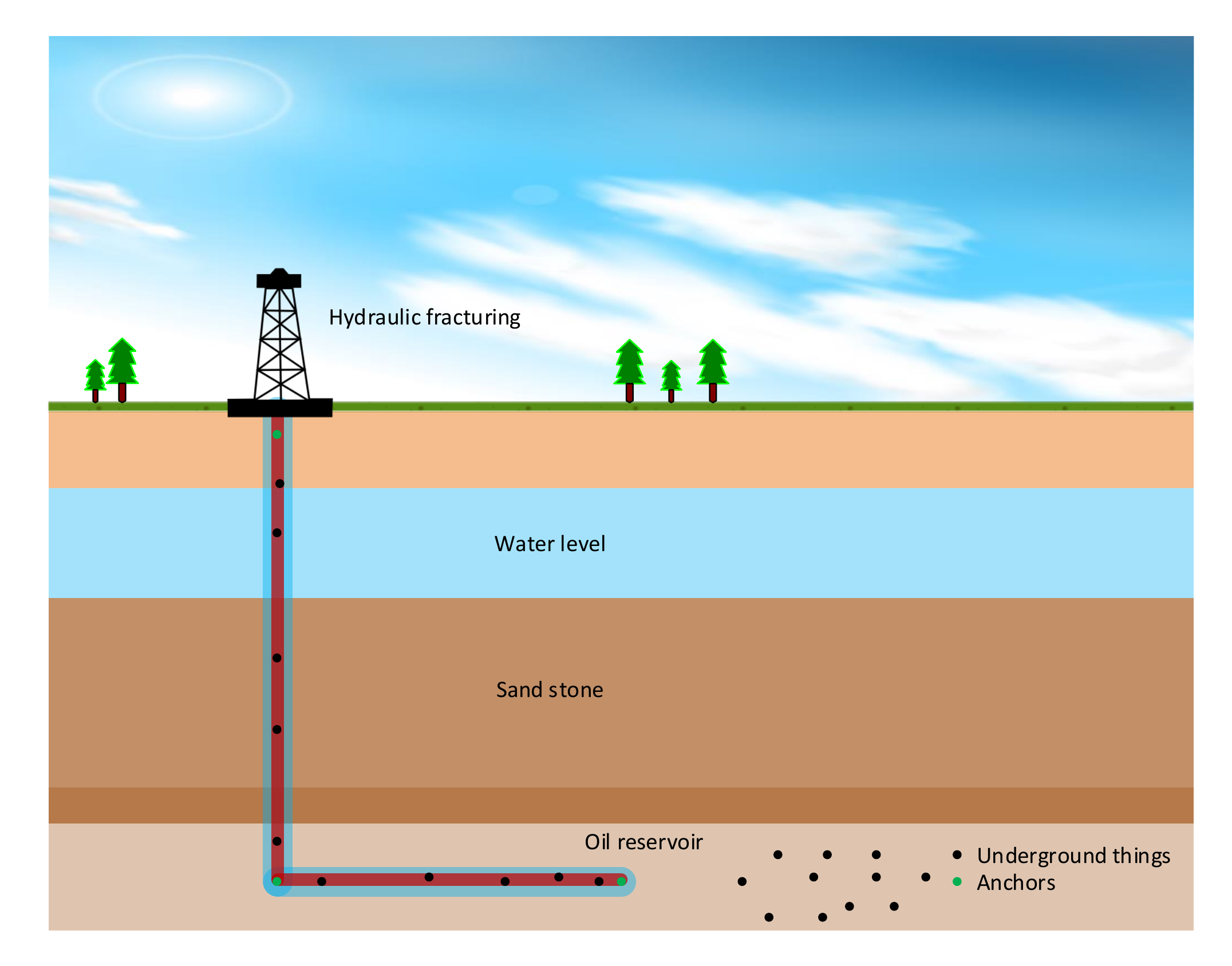}  
\caption{System model.\label{fig:underground}} 
\vspace{-1.5 em}
\end{figure}

However, in sparse underground sensing networks the connectivity of the network is limited due to the short transmission distance of MI communication. Hence, localization of the underground things is challenging due to the limited connectivity, directionality of MI coils, interference from the Earth’s magnetic field and underground metals \cite{ Markham2012, huang2018, Yan2018}. Hence, in \cite{Abrudan2016} the authors have investigated one of the challenges mentioned above, i.e., the impact of minerals and rocks on the MI-based underground localization. Attenuation properties have been estimated for different mediums in the underground which significantly affects the performance of the localization techniques.

All of the above works do not characterize the achievable accuracy of the developed MI-based localization techniques for underground sensing networks which is a crucial and challenging task. In the past, the estimation bounds for the time of arrival \cite{Imtiaz2016,Ansari2018}, the angle of arrival \cite{Duan2015}, time difference of arrival \cite{Dersan2002}, and received signal strength \cite{Ciftler2015} based localization techniques have been investigated. Subsequently, these findings have opened the door for developing accurate and robust localization algorithms. Therefore, we expect the same for our proposed lower bound for MI-based IOUT. The performance of the localization techniques is characterized by the CRLB which is a non-linear estimation problem. Different CRLB analysis exists in the literature which not only depends on the ranging method but also depend on the other parameters such as multipath effect, number of anchors, and network type (single hop or multi-hop) \cite{patwari2003relative,Jia2008,saeed64cluster}.  Due to the simplicity and generic expressions of the CRLB, it is an attractive analyzing tool for localization systems.  Therefore, in this paper, we derive the expression of the CRLB for MI-based IOUT localization which takes into account the channel parameters of the underground magnetic-induction.  The derived CRLB provide the suggestions for an MI-based underground localization system by associating the system parameters with the error trend.

\section{System Model}
In this section, we first introduce the 3D architecture for MI-based IOUT. Then the wireless propagation model is presented for the distance estimation. 
\label{systmodel}

\subsection{MI-based IOUT Setup}
We consider the conventional network setup for 3D MI-based IOUT which consists on $N$ number of underground things and $M$ number of anchor nodes as shown in Fig.~\ref{fig:underground}. The underground things are injected into the oil reservoir by using hydraulic fracturing \cite{Guo2014, Akkaş2017}.  The underground things are uniformly distributed, and their positions are denoted by $\{\boldsymbol{s}_i\}_{i=1}^N$ where $\boldsymbol{s}_i =\{{x_i,y_i,z_i}\}$ represents the 3D position of the $i$-th underground thing. As anchor nodes are necessary to find the position of the underground things, we assume that the anchor nodes are attached to the fracturing well with known positions $\{\boldsymbol{s}_j\}_{j=1}^M$ where $\boldsymbol{s}_j = \{{x_j,y_j,z_j}\}$ is the 3D position of the $j$-th anchor. The anchor nodes use large dipole antennas to communicate with the underground things by using the MI communication link. Hence, the downlink is a single hop channel while the underground things use multi-hop channel for the uplink transmission due to their limited range. We also assume that the anchor nodes have higher transmission range as they can be attached to the external power sources \cite{Guo2014}. The underground things can communicate to close by underground things and the anchors by using magnetic induction.  Based on the above network setup the problem of localization is defined as, to estimate the unknown location of underground things for a given set of anchors and estimated MI-based distances.

\begin{figure}
\centering
\includegraphics[width=1\columnwidth]{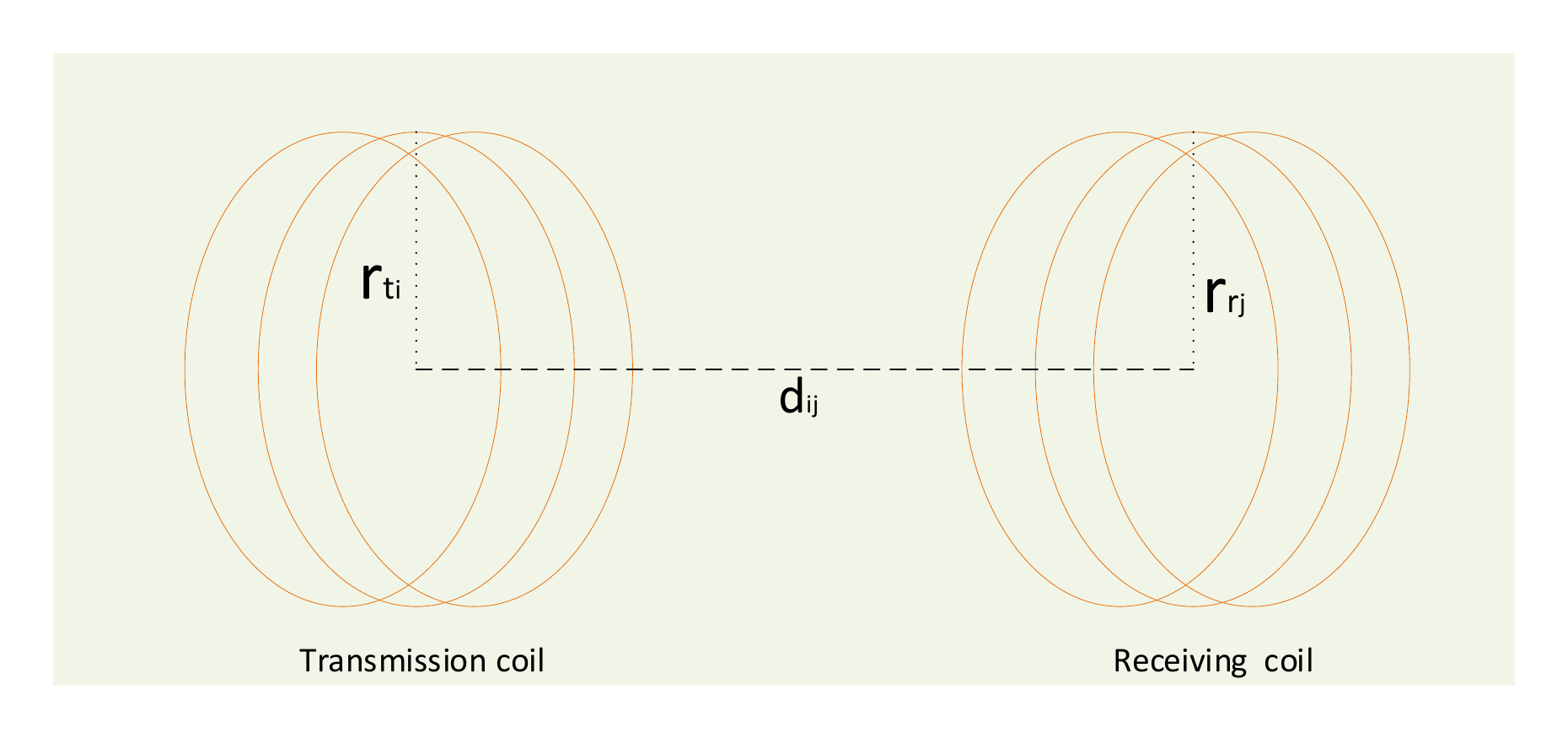}  
\caption{MI communication link.\label{fig:micoils}} 
\vspace{-1.5 em}
\end{figure}
\subsection{MI-based Underground Distance Estimation}
\begin{figure}
\centering
\includegraphics[width=1\columnwidth]{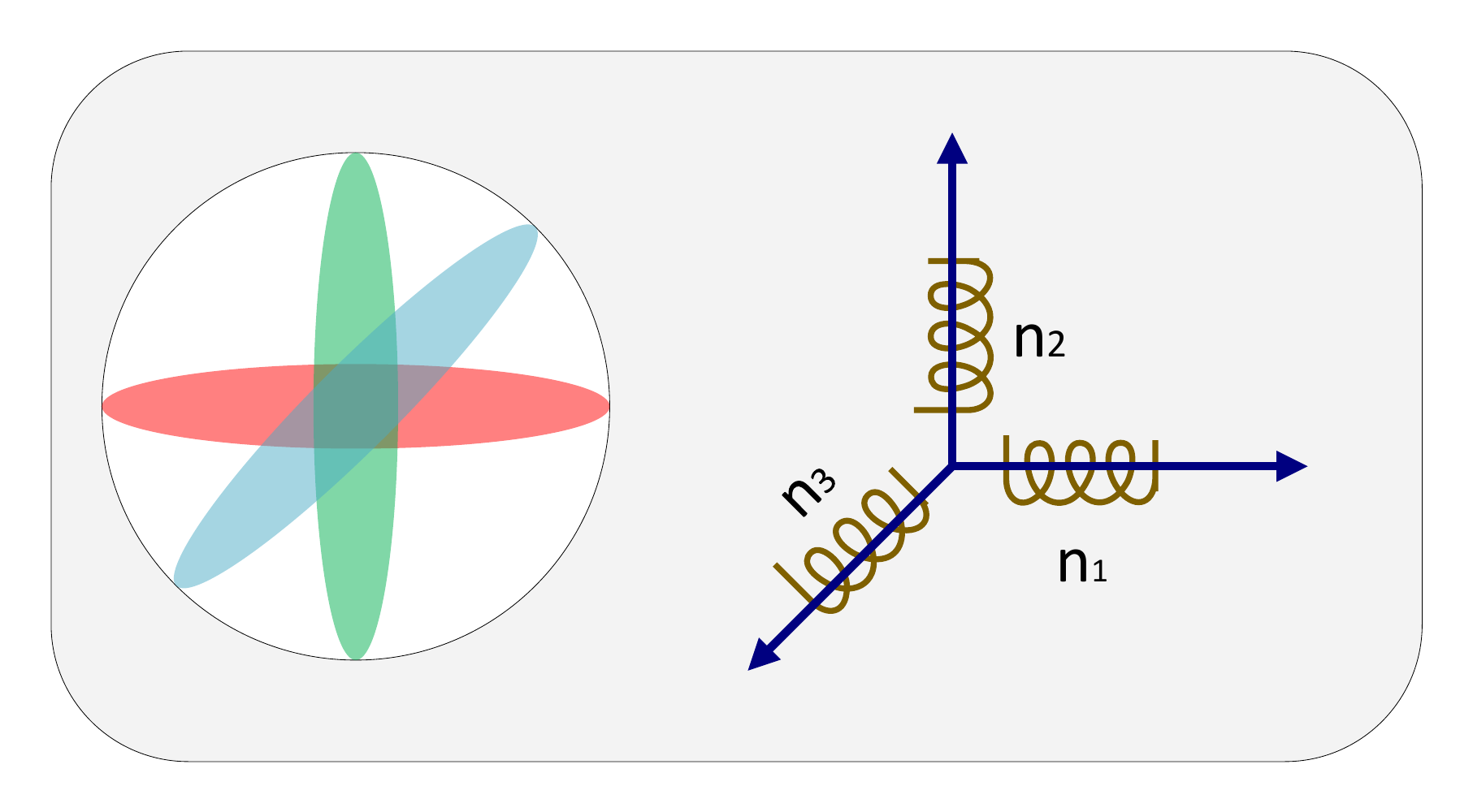}  
\caption{Tri-directional MI communication\label{fig:trimicoil}} 
\vspace{-1.5 em}
\end{figure}

The information exchange between the transmitting node and the receiving node in MI-based IOUT is accomplished by using a time-varying magnetic field which is produced by the modulated sinusoidal signal from the transmitter coil antenna. The time-varying magnetic field induces current at the receiver coil antenna which is demodulated to retrieve the information. Fig.~\ref{fig:micoils} shows a realization of an MI-based transceiver. Consider that the current in the transmitting coil is $I = I_0 e^{-j\omega t}$, where $I_0$ is the direct current, $\omega$ is the angular frequency, and $t$ is the instantaneous time. This current can then induce current in the nearby coil by the phenomena of mutual induction. However, a single coil cannot guarantee to receive an MI signal if the receiving coil is not well coupled with the transmitting coil. Therefore, in the harsh underground environment, we assume the tri-directional coil receiver structure proposed in \cite{Tan2015} for receiving strong MI signal (see Fig \ref{fig:trimicoil}). Based on the magnetic induction, the relationship between the transmit and received power at high frequency $f$ and large number of transmitter coil turns $N_t$ is given in \cite{Sun2009} as
\begin{equation}\label{eq: pr}
P_{r_j} = \frac{\omega \mu P_{t_i} N_{r_j} r_{t_i}^3 r_{r_j}^3\sin^2\alpha_{ij}}{16 R_0 d_{ij}^6},
\end{equation}
where $\omega$ is the angular frequency, $\mu$ is the permeability of soil, $P_{t_i}$ is the transmit power, $N_{r_j}$ is the number of turns in the receiver coil, $r_{t_i}$ and $r_{r_i}$ are the diameters of the transmitter and receiver coil respectively, $\alpha_{ij}$ is the angle between the axes of the transmitter and receiver coils, $R_0$ is the resistance of a unit length loop, and $d_{ij} = \parallel \boldsymbol{s}_i - \boldsymbol{s}_j \parallel$ is the distance between the transmitter and the receiver coil. It is worthy to note that the received power expression in \eqref{eq: pr}  has been experimentally validated in \cite{Tan2015}. Additionally, the received power is affected by background noise $b_{ij}$ which is modelled as zero mean Gaussian random variable  with variance $\sigma_{ij}^2$ \cite{Tan2015}. The  probability density function (PDF) of the noisy received power $\tilde{P}_{r_j}$ is written as
\begin{equation}\label{eq: pdf}
f(\tilde{P}_{r_j}|\boldsymbol{s}_i,\boldsymbol{s}_j) = \frac{1}{\sigma_{ij}\sqrt{2 \pi}}\exp^{\bigg(-\frac{(\tilde{P}_{r_j}-{P}_{r_j})^2}{2\sigma_{ij}^2}\bigg)}.
\end{equation}

\section{Achievable Accuracy of MI-based IOUT Localization}
The achievable accuracy of any localization technique can be characterized by using the unbiased estimator, i.e., the CRLB. Therefore, in this section we derive the closed form expression for the MI-based IOUT localization. The derivation of the CRLB consists of the following steps.

\subsubsection{Log-likelihood Ratio Calculation}  The log-likelihood ratio is calculated from the PDF in \eqref{eq: pdf} which is written as
\begin{equation} \label{eq: likelihoodratio}
\ell[dB] = -\log(\sigma_{ij}\sqrt{2\pi})+\log\Bigg(\exp^{-\frac{(\tilde{P}_{r_j}-{P}_{r_j})^2}{2\sigma_{ij}^2}}\Bigg).
\end{equation}
The joint log-likelihood ratio for all the underground things is given as
\begin{equation}\label{eq: likelihood}
\boldsymbol{L} = \sum_{i=1}^N\sum_{j=1}^M \log \bigg( f(\tilde{P}_{r_j}|\boldsymbol{s}_i,\boldsymbol{s}_j)\bigg).
\end{equation}
\subsubsection{Computation of the Fisher Information Matrix} Fisher information matrix (FIM) tells us about how much information (localization accuracy) can be achieved from the noisy received power \cite{Pakrooh2015}. The FIM constitutes of sub-matrices for the three-dimensional location estimation given as
\begin{equation} \label{eq: fisher}
\boldsymbol{I} = \begin{bmatrix} \boldsymbol{I}_{x,x} & \boldsymbol{I}_{x,y} & \boldsymbol{I}_{x,z}\\
\boldsymbol{I}_{x,y}^T & \boldsymbol{I}_{y,y} & \boldsymbol{I}_{y,z} \\
\boldsymbol{I}_{x,z}^T & \boldsymbol{I}_{y,z}^T & \boldsymbol{I}_{z,z} \\
\end{bmatrix}.
\end{equation}
The elements of the sub-matrices in FIM are derived from the second order derivatives of the log-likelihood function defined in \eqref{eq: likelihood}. The elements of the sub-matrices are calculated as
\begin{equation} \label{eq: f1}
\boldsymbol{I}_{x,x_{i=l}} = 
E \bigg(\frac{\partial^2 \ell_{ij}}{\partial x_i^2}\bigg),
\end{equation}
\begin{equation}
\boldsymbol{I}_{x,y_{i=l}} = 
E \bigg(\frac{\partial^2 \ell_{ij}}{\partial x_i y_i}\bigg),
\end{equation}
\begin{equation}
\boldsymbol{I}_{x,z_{i=l}} = 
E \bigg(\frac{\partial^2 \ell_{ij}}{\partial x_i z_i}\bigg),
\end{equation}
\begin{equation}
\boldsymbol{I}_{y,y_{i=l}} = 
E \bigg(\frac{\partial^2 \ell_{ij}}{\partial y_i^2}\bigg),
\end{equation}
\begin{equation}
\boldsymbol{I}_{y,z_{i=l}} = 
E \bigg(\frac{\partial^2 \ell_{ij}}{\partial y_i z_i}\bigg),
\end{equation}
and
\begin{equation} \label{eq: ff}
\boldsymbol{I}_{z,z_{i=l}} = 
E \bigg(\frac{\partial^2 \ell_{ij}}{\partial z_i^2}\bigg),
\end{equation}
respectively. The non-diagonal elements are given as
\begin{equation} \label{eq: fnon1}
\boldsymbol{I}_{x,x_{i \neq l}} = 
E \bigg(\frac{\partial^2 \ell_{ij}}{\partial x_i x_l}\bigg),
\end{equation}
\begin{equation}
\boldsymbol{I}_{x,y_{i \neq l}} = 
E \bigg(\frac{\partial^2 \ell_{ij}}{\partial x_i y_l}\bigg),
\end{equation}
\begin{equation}
\boldsymbol{I}_{x,z_{i \neq l}} = 
E \bigg(\frac{\partial^2 \ell_{ij}}{\partial x_i z_l}\bigg),
\end{equation}
\begin{equation}
\boldsymbol{I}_{y,y_{i \neq l}} = 
E \bigg(\frac{\partial^2 \ell_{ij}}{\partial y_i y_l}\bigg),
\end{equation}
\begin{equation}
\boldsymbol{I}_{y,z_{i\neq l}} = 
E \bigg(\frac{\partial^2 \ell_{ij}}{\partial y_i z_l}\bigg),
\end{equation}
and
\begin{equation} \label{eq: fnonff}
\boldsymbol{I}_{z,z_{i\neq l}} = 
E \bigg(\frac{\partial^2 \ell_{ij}}{\partial z_i z_l}\bigg),
\end{equation}
where $E(\cdot)$ is the expectation operator. To derive the diagonal elements of each sub-matrix, we put the value of $P_{r_j}$ in \eqref{eq: f1} to \eqref{eq: ff} and solve it. Expression for the diagonal elements are obtained in Appendix A as follows
\begin{eqnarray} \label{eq: ffd1}
\boldsymbol{I}_{x,x_{i=l}} &=& 
\frac{3k}{\sigma_{ij}^2}\Bigg(\frac{2k}{\parallel \boldsymbol{s}_i - \boldsymbol{s}_j\parallel^7}-\frac{28k (x_i - x_j)^2}{\parallel \boldsymbol{s}_i - \boldsymbol{s}_j\parallel^8} \Bigg. \nonumber \\
& & \Bigg. +\frac{k}{\parallel \boldsymbol{s}_i - \boldsymbol{s}_j\parallel^7}- 
\frac{8(x_i-x_j)^2)}{\parallel \boldsymbol{s}_i - \boldsymbol{s}_j\parallel^5}
\Bigg)
\end{eqnarray}
\begin{equation}
\boldsymbol{I}_{x,y_{i=l}} = 
\frac{60 k^2 (x_i - x_j)(y_i - y_j)}{\sigma_{ij}^2\parallel \boldsymbol{s}_i - \boldsymbol{s}_j\parallel^8},
\end{equation}
\begin{equation}
\boldsymbol{I}_{x,z_{i=l}} = 
\frac{60 k^2 (x_i - x_j)(z_i - z_j)}{\sigma_{ij}^2\parallel \boldsymbol{s}_i - \boldsymbol{s}_j\parallel^8},
\end{equation}
\begin{eqnarray} 
\boldsymbol{I}_{y,y_{i=l}} &=& 
\frac{3k}{\sigma_{ij}^2}\Bigg(\frac{2k}{\parallel \boldsymbol{s}_i - \boldsymbol{s}_j\parallel^7}-\frac{28k (y_i - y_j)^2}{\parallel \boldsymbol{s}_i - \boldsymbol{s}_j\parallel^8} \Bigg. \nonumber \\
& & \Bigg. +\frac{k}{\parallel \boldsymbol{s}_i - \boldsymbol{s}_j\parallel^7}- 
\frac{8(y_i-y_j)^2)}{\parallel \boldsymbol{s}_i - \boldsymbol{s}_j\parallel^5}
\Bigg)
\end{eqnarray}
\begin{equation}
\boldsymbol{I}_{y,z_{i=l}} = 
\frac{60 k^2 (y_i - y_j)(z_i - z_j)}{\sigma_{ij}^2\parallel \boldsymbol{s}_i - \boldsymbol{s}_j\parallel^8},
\end{equation}
and
\begin{eqnarray} 
\boldsymbol{I}_{z,z_{i=l}} &=& 
\frac{3k}{\sigma_{ij}^2}\Bigg(\frac{2k}{\parallel \boldsymbol{s}_i - \boldsymbol{s}_j\parallel^7}-\frac{28k (z_i - z_j)^2}{\parallel \boldsymbol{s}_i - \boldsymbol{s}_j\parallel^8} \Bigg. \nonumber \\
& & \Bigg. +\frac{k}{\parallel \boldsymbol{s}_i - \boldsymbol{s}_j\parallel^7}- 
\frac{8(z_i-z_j)^2)}{\parallel \boldsymbol{s}_i - \boldsymbol{s}_j\parallel^5}
\Bigg)
\end{eqnarray}
where $k = \frac{\omega \mu P_{t_i} N_{r_j} r_{t_i}^3 r_{r_j}^3\sin^2\alpha_{ij}}{16 R_0}$. Similarly solving \eqref{eq: fnon1} to \eqref{eq: fnonff} for the non-diagonal elements yields
\begin{eqnarray} \label{eq: fx}
\boldsymbol{I}_{x,x_{i \neq l}} &=& 
-\frac{3k}{\sigma_{ij}^2}\Bigg(\frac{2k}{\parallel \boldsymbol{s}_i - \boldsymbol{s}_j\parallel^7}-\frac{28k (x_i - x_j)^2}{\parallel \boldsymbol{s}_i - \boldsymbol{s}_j\parallel^8} \Bigg. \nonumber \\
& & \Bigg. +\frac{k}{\parallel \boldsymbol{s}_i - \boldsymbol{s}_j\parallel^7}- 
\frac{8(x_i-x_j)^2)}{\parallel \boldsymbol{s}_i - \boldsymbol{s}_j\parallel^5}
\Bigg)
\end{eqnarray}
\begin{equation}
\boldsymbol{I}_{x,y_{i \neq l}} = 
-\frac{60 k^2 (x_i - x_j)(y_i - y_j)}{\sigma_{ij}^2\parallel \boldsymbol{s}_i - \boldsymbol{s}_j\parallel^8},
\end{equation}
\begin{equation}
\boldsymbol{I}_{x,z_{i\neq l}} = 
-\frac{60 k^2 (x_i - x_j)(z_i - z_j)}{\sigma_{ij}^2\parallel \boldsymbol{s}_i - \boldsymbol{s}_j\parallel^8},
\end{equation}
\begin{eqnarray} 
\boldsymbol{I}_{y,y_{i\neq l}} &=& 
-\frac{3k}{\sigma_{ij}^2}\Bigg(\frac{2k}{\parallel \boldsymbol{s}_i - \boldsymbol{s}_j\parallel^7}-\frac{28k (y_i - y_j)^2}{\parallel \boldsymbol{s}_i - \boldsymbol{s}_j\parallel^8} \Bigg. \nonumber \\
& & \Bigg. +\frac{k}{\parallel \boldsymbol{s}_i - \boldsymbol{s}_j\parallel^7}- 
\frac{8(y_i-y_j)^2)}{\parallel \boldsymbol{s}_i - \boldsymbol{s}_j\parallel^5}
\Bigg)
\end{eqnarray}
\begin{equation}
\boldsymbol{I}_{y,z_{i\neq l}} = 
-\frac{60 k^2 (y_i - y_j)(z_i - z_j)}{\sigma_{ij}^2\parallel \boldsymbol{s}_i - \boldsymbol{s}_j\parallel^8},
\end{equation}
and
\begin{eqnarray}\label{eq: ffdf}
\boldsymbol{I}_{z,z_{i\neq l}} &=& 
-\frac{3k}{\sigma_{ij}^2}\Bigg(\frac{2k}{\parallel \boldsymbol{s}_i - \boldsymbol{s}_j\parallel^7}-\frac{28k (z_i - z_j)^2}{\parallel \boldsymbol{s}_i - \boldsymbol{s}_j\parallel^8} \Bigg. \nonumber \\
& & \Bigg. +\frac{k}{\parallel \boldsymbol{s}_i - \boldsymbol{s}_j\parallel^7}- 
\frac{8(z_i-z_j)^2)}{\parallel \boldsymbol{s}_i - \boldsymbol{s}_j\parallel^5}
\Bigg).
\end{eqnarray}
Substituting the values of \eqref{eq: ffd1} to \eqref{eq: ffdf} in \eqref{eq: fisher} completes the FIM.

\subsubsection{Cramer Rao Lower Bound} Finally the CRLB is calculated from the inverse of the FIM which is given as
\begin{equation}
\text{CRLB} = \boldsymbol{I}_{x,x}^{-1}+\boldsymbol{I}_{y,y}^{-1}+\boldsymbol{I}_{z,z}^{-1}.
\end{equation}
Thus, we have developed a generalized CRLB for MI-based IOUT which is the function of different channel and network parameters such as operating frequency, number of turns of the coils, radius of the coils, transmit power, noise variance, number of anchors, and number of underground things. The derived CRLB provide the suggestions for an MI-based underground localization system by associating the channel and network parameters with the error trend.
\section{Numerical Results}
In this section, we provide numerical results to validate the derived CRLB in a practical 3D oil reservoir setup. The performance of the CRLB for MI-based  IOUT is tested under various network settings. Table \ref{Tableparameters} presents the simulation parameters which are taken mainly from \cite{Lin2017}. In the following, first, we examine the effect of operating frequency, noise variance, and the number of anchors on the performance of the CRLB. Then, we evaluate the performance regarding the number of turns of the coils, coils size, and transmit power.
\begin{table}[htb!]
\footnotesize
\centering
\caption{Simulation parameters}
\label{Tableparameters}
\begin{tabular}{| p{4.5cm} | p{3.5cm}|}
\hline
\hline
 Parameters              & Values     \\ \hline
 Operating frequency   & 7 MHz and 13 MHz         \\
 Fracture area            & 8 $\times$ 8 $\text{m}^2$ \\
 Depth of the fracture   & 1.8 Km \\
 Coil radius                   & 0.01-0.04 m \\
 Number of turns in coils          & 10-30               \\
 Transmit power                    & 100-200 mW                       \\
 Unit length resistance of antenna               &  0.01 $\Omega/m$         \\
 Temperature            & 418 K              \\
 Noise variance                 & 0.05-0.7 m\\
 Number of underground things & 60\\
\hline
\hline        
\end{tabular}
\end{table}
\begin{figure}
\centering
\includegraphics[width=0.8\columnwidth]{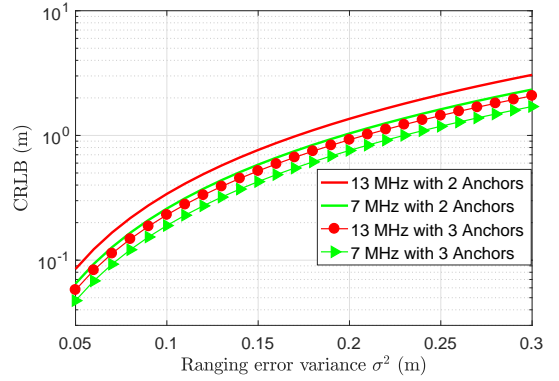}  
\caption{CRLB vs. Ranging error variance and frequency\label{fig:frequency}} 
\end{figure}
Fig. \ref{fig:frequency} shows the adverse effect of ranging error variance on the accuracy of CRLB for IOUT. Additionally, we have also demonstrated the impact of changing the operating frequency where two different frequencies of 7 MHz and 13 MHz are used. It is clear from Fig. \ref{fig:frequency} that increasing the frequency results in low accuracy due to high path loss.
Moreover, Fig. \ref{fig:frequency} also shows that increasing the number of anchors from 2 to 3 improves the accuracy of the CRLB. Note that the results in Fig. \ref{fig:frequency} are averaged over 500 different network setups with 60 underground things randomly distributed in 8 $\times$ 8 $m^2$ fracture area.

\begin{figure}
\centering
\includegraphics[width=0.8\columnwidth]{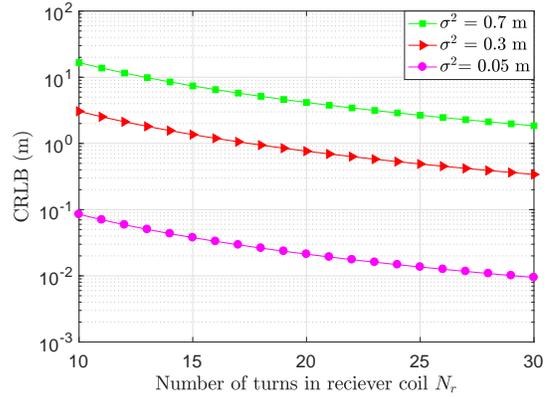}  
\caption{CRLB vs. Number of turns in the receiver\label{fig:turns}} 
\end{figure}

\begin{figure}
\centering
\includegraphics[width=0.8\columnwidth]{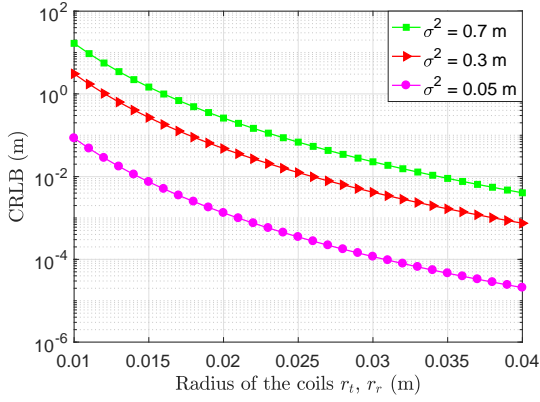}  
\caption{CRLB vs. Coil size\label{fig:coilsize}} 
\end{figure}

Fig. \ref{fig:turns} shows CRLB as a function of the number of turns in the coil of the underground sensors. Results of CRLB in Fig. \ref{fig:turns} are obtained at a frequency of 13 MHz with noise variances of 0.05, 0.3, and 0.7 m respectively. Fig. \ref{fig:turns} shows that increasing the number of turns of the coils improve the localization accuracy. However, increasing the number of turns of the coil may increase the size of the coil where the small size of fracturing well requires small size coils. Therefore, in Fig. \ref{fig:coilsize}, we have investigated the impact of coil size on the performance of the CRLB. It is clear from  Fig. \ref{fig:coilsize} that for a given ranging error variance the accuracy of the CRLB improves with the increase in the coil size from 0.01 m to 0.04 m. It is worthy to note here that the coil design parameters such as the number of turns and coil size need to be adjusted based on the results in Fig. \ref{fig:turns} and Fig. \ref{fig:coilsize} to achieve a certain localization accuracy.

\begin{figure}
\centering
\includegraphics[width=0.8\columnwidth]{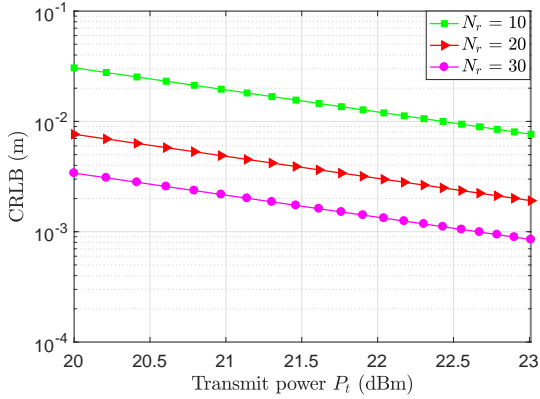}  
\caption{CRLB vs. Transmit power\label{fig:power}} 
\vspace{-1.5 em}
\end{figure}

Finally, we evaluate the performance of the derived CRLB for MI-based IOUT regarding the transmit power. The transmit power of commercially available coils for MI communications varies from 100 mW (20 dBm) to 200 mW (23 dBm). Therefore, we kept the range of 20-23 dBm for the transmit power with the different type of coils. Fig. \ref{fig:power} shows that increasing the transmit power improves the accuracy of the CRLB for a given type of coil.

According to the above results, the CRLB is the function of MI-based channel and network parameters. Therefore, all these parameters should be taken into account to develop a robust and accurate MI-based IOUT localization system.

\section{Conclusions}
In this paper, we derived the CRLB for three-dimensional localization of MI-based IOUT. We examined the effects of numerous parameters such as the number of anchors, operating frequency, coil size, number of turns in the coil, ranging error, and transmit power on the derived lower bound. Subsequently, we found out from our analysis that increasing the number of anchors improves the localization accuracy. Moreover, increasing the operating frequency increases the lower bound. Conversely, increasing the number of turns in the coil or reducing the size of the coil reduces the lower bound. Furthermore, the effect of transmitting power is evaluated which shows that an increase in transmit power reduces the CRLB.
The analysis and findings of this paper should open the door for designing efficient localization algorithms for MI-based IOUT.

\appendices
\section{Derivation of the Elements for the FIM}
Using the PDF in \eqref{eq: likelihoodratio} the partial derivatives are calculated as
\begin{equation} \label{eq: d1}
\frac{\partial \ell_{ij}}{\partial x_i}= \frac{1}{2 \sigma_{ij}^2}\bigg(-\frac{\partial{\hat{P}}_{r_j}^2}{\partial x_i}-\frac{\partial {P}_{r_j}^2}{\partial x_i}+ \frac{\partial {P}_{r_j} {\hat{P}}_{r_j}}{\partial x_i}\bigg).
\end{equation}

Putting the value of ${P}_{r_j}$ in \eqref{eq: d1} yields
\begin{eqnarray} \label{eq: d2}
\frac{\partial \ell_{ij}}{\partial x_i} &=& \frac{1}{2 \sigma_{ij}^2} \bigg( -\frac{\partial}{\partial x_i}\left( k^2 {\parallel \boldsymbol{s}_i - \boldsymbol{s}_j\parallel^{-6}}\right)
 \bigg. \nonumber \\
 & & \bigg. + {\hat{P}}_{r_j} \frac{\partial}{\partial x_i} \left( k {\parallel \boldsymbol{s}_i - \boldsymbol{s}_j\parallel^{-3}}\right)  \bigg).
\end{eqnarray}
Taking the partial derivative of \eqref{eq: d2} with respect to $x_i$ yields

\begin{eqnarray} \label{eq: d3}
\frac{\partial \ell_{ij}}{\partial x_i} = \frac{1}{2 \sigma_{ij}^2} \bigg( \frac{ k^2 12 (x_i -x_j)}{{\parallel \boldsymbol{s}_i - \boldsymbol{s}_j\parallel^{7}}}
 +   \frac{ k {\hat{P}}_{r_j} 6 (x_i -x_j)}{{\parallel \boldsymbol{s}_i - \boldsymbol{s}_j\parallel^{6}}}\bigg).
\end{eqnarray}

Now the second-order partial derivative of \eqref{eq: d3} results in

\begin{eqnarray} \label{eq: d3}
\frac{\partial^2 \ell_{ij}}{\partial x_i^2}&=& \frac{k}{\sigma_{ij}^2}\bigg(\frac{6 k }{{\parallel \boldsymbol{s}_i - \boldsymbol{s}_j\parallel^{7}}} - \frac{84 k (x_i -x_j)^2 }{{\parallel \boldsymbol{s}_i - \boldsymbol{s}_j\parallel^{8}}}  \bigg. \nonumber \\
& & \bigg. + \frac{3 {\hat{P}}_{r_j} }{{\parallel \boldsymbol{s}_i - \boldsymbol{s}_j\parallel^{4}}} - \frac{24 (x_i -x_j)^2 }{{\parallel \boldsymbol{s}_i - \boldsymbol{s}_j\parallel^{5}}}
\bigg).
\end{eqnarray}
Now using the fact that $E({\hat{P}}_{r_j}) = {P}_{r_j}$, \eqref{eq: d3} is simplified as

\begin{eqnarray} \label{eq: d4}
E\bigg(\frac{\partial^2 \ell_{ij}}{\partial x_i^2}\bigg)&=& \frac{3 k}{\sigma_{ij}^2}\bigg(\frac{2 k }{{\parallel \boldsymbol{s}_i - \boldsymbol{s}_j\parallel^{7}}} - \frac{28 k (x_i -x_j)^2 }{{\parallel \boldsymbol{s}_i - \boldsymbol{s}_j\parallel^{8}}}  \bigg. \nonumber \\
& & \bigg. + \frac{k }{{\parallel \boldsymbol{s}_i - \boldsymbol{s}_j\parallel^{7}}} - \frac{8 (x_i -x_j)^2 }{{\parallel \boldsymbol{s}_i - \boldsymbol{s}_j\parallel^{8}}}
\bigg).
\end{eqnarray}

Similarly, the other terms can be easily obtained as follows

\begin{eqnarray}
E\bigg(\frac{\partial^2 \ell_{ij}}{\partial y_i^2}\bigg) &=& \frac{3k}{\sigma_{ij}^2}\Bigg(\frac{2k}{\parallel \boldsymbol{s}_i - \boldsymbol{s}_j\parallel^7}-\frac{28k (y_i - y_j)^2}{\parallel \boldsymbol{s}_i - \boldsymbol{s}_j\parallel^8} \Bigg. \nonumber \\
& & \Bigg. +\frac{k}{\parallel \boldsymbol{s}_i - \boldsymbol{s}_j\parallel^7}- 
\frac{8(y_i-y_j)^2)}{\parallel \boldsymbol{s}_i - \boldsymbol{s}_j\parallel^5}
\Bigg),
\end{eqnarray}

\begin{eqnarray}
E\bigg(\frac{\partial^2 \ell_{ij}}{\partial z_i^2}\bigg) &=& \frac{3k}{\sigma_{ij}^2}\Bigg(\frac{2k}{\parallel \boldsymbol{s}_i - \boldsymbol{s}_j\parallel^7}-\frac{28k (z_i - z_j)^2}{\parallel \boldsymbol{s}_i - \boldsymbol{s}_j\parallel^8} \Bigg. \nonumber \\
& & \Bigg. +\frac{k}{\parallel \boldsymbol{s}_i - \boldsymbol{s}_j\parallel^7}- 
\frac{8(z_i-z_j)^2)}{\parallel \boldsymbol{s}_i - \boldsymbol{s}_j\parallel^5}
\Bigg)
\end{eqnarray}

\begin{equation}
E \bigg(\frac{\partial^2 \ell_{ij}}{\partial x_i y_i}\bigg) = 
\frac{60 k^2 (x_i - x_j)(y_i - y_j)}{\sigma_{ij}^2\parallel \boldsymbol{s}_i - \boldsymbol{s}_j\parallel^8},
\end{equation}

\begin{equation}
E \bigg(\frac{\partial^2 \ell_{ij}}{\partial x_i z_i}\bigg) = 
\frac{60 k^2 (x_i - x_j)(z_i - z_j)}{\sigma_{ij}^2\parallel \boldsymbol{s}_i - \boldsymbol{s}_j\parallel^8},
\end{equation}

\begin{equation}
E \bigg(\frac{\partial^2 \ell_{ij}}{\partial y_i z_i}\bigg) = 
\frac{60 k^2 (y_i - y_j)(z_i - z_j)}{\sigma_{ij}^2\parallel \boldsymbol{s}_i - \boldsymbol{s}_j\parallel^8}.
\end{equation}
The non-diagonal terms can also be obtained in the similar fashion. Finally all these derived elements complete the FIM for the CLRB.
\bibliographystyle{../bib/IEEEtran}
\bibliography{../bib/IEEEabrv,../bib/nasir_ref}

\end{document}